\documentclass[a4paper, 12pt]{article} 
\usepackage{graphicx}      
\usepackage{float}	
\usepackage{tikz}
\usepackage{amsmath}
\usepackage{amssymb}
\usepackage{subfig}
\usepackage{dblfloatfix}
\usepackage{scalefnt}
\usepackage[linesnumbered, ruled, vlined]{algorithm2e}
\usepackage[letterpaper,width=176mm,top=20.1mm,bottom=15.4mm]{geometry}
\newtheorem{remark}{Remark}

\title{\Large \bf
Data-driven control of infinite dimensional systems: \\ Application to a continuous crystallizer
}

\author{Pauline Kergus$^{1}$ 
\thanks{$^{1}$Automatic Control Department, Lund University, Sweden}%
\thanks{Contact: \tt \small{pauline.kergus@control.lth.se}}
\thanks{The author is funded by the European Research Council (ERC) under the European Union's Horizon 2020 research and innovation program under grant agreement No 834142 (ScalableControl). She is also a member of the ELLIIT Strategic Research Area at Lund University.}
}

\begin{document}
\scalefont{.96}
\maketitle
\thispagestyle{empty}
\pagestyle{empty}

\tikzstyle{sum}=[draw,circle,text width = 0.3cm]
\tikzstyle{block}=[rectangle, draw=black, text centered]

\newcommand*{\new}{\textcolor{black}}

\begin{abstract}
Controlling infinite dimensional models remains a challenging task for many practitioners since they are not suitable for traditional control design techniques or will result in a high-order controller too complex for implementation. Therefore, the model or the controller need to be reduced to an acceptable dimension, which is time-consuming, requires some expertise and may introduce numerical error. This paper tackles the control of such a system, namely a continuous crystallizer, and compares two different data-driven  strategies: the first one is a structured robust technique while the other one, called \textbf{L-DDC}, is based on the Loewner interpolatory framework.
\end{abstract}

\section{Introduction}
In general, control problems can be divided into four categories according to the available description of the process to be controlled, see \cite{hou2013model}: i) those for which an accurate model is available, ii) those where the system is described by a roughly accurate model with moderate uncertainties, iii) those described by models that are too complex for traditional model-based design techniques and iv) those for which models are not available or hard to obtain. While the first category can be covered by traditional model-based control techniques, the second one motivated the development of specific techniques such as robust or adaptive control which, in some cases, have been extended to tackle problems from the third category. Finally, the development of data-driven control techniques has been motivated by the third and fourth categories.

Infinite dimensional systems are representative of the problems that can be found in the third category. Due to their complexity, they are in most cases not suitable for model-based techniques. However, when a transfer function is available (for linear PDEs for instance), robust control theory has been extended to the infinite-dimensional case \cite{foias1996robust}. As explained in \cite{morris2010control}, the main drawback of this approach is that the resulting controller is also of infinite dimension and needs to be reduced. In practice, a reduced model is obtained from the infinite-dimensional transfer function and traditional model-based techniques are then used to design a controller. To sum up, both approaches require to reduce either the controller or the model, which is time-consuming and requires additional expertise. In addition, in the first case, the reduced controller may not reach the level of performance it was meant to achieve and the guarantees furnished by robust control may not apply anymore. On the other side, when reducing the infinite-dimensional system to a finite order one in order to apply traditional model-based techniques, the resulting controller is not tailored for the actual system but for the reduced model. The Loewner Data-Driven Control (\textbf{L-DDC}), detailed in \cite{kergus2019}, allows to obtain a reduced-order controller tailored to the actual system on the basis of samples from its frequency response, which, for this category of systems, can be extracted directly from the transfer function. This data-driven approach then constitutes another way to use order reduction to control infinite dimensional systems. On the other side, the data-driven structured $\mathcal{H}_\infty$ design technique introduced in \cite{apkarian2017structured} has been proposed to tackle the control of infinite dimensional systems without using any order reduction technique.

The objective of this paper is two-fold. The first one is to highlight the relevance of the data-driven approaches in \cite{kergus2019} and \cite{apkarian2017structured} to tackle the control of infinite dimensional systems where the transfer function is available. The second objective of this paper is to compare these two approaches, based on their application to the control of a continuous crystallizer, described by an infinite-dimensional transfer function, as detailed in \cite{rachah2016mathematical}. This problem has already been tackled in \cite{apkarian2017structured} but also in \cite{vollmer2001h} using $H_\infty$ control directly on the infinite-dimensional model (model-based approach).

The rest of this paper is organized as follows. Section \ref{sec:cry} introduces the considered application, the continuous crystallizer, and sums up two control strategies previously used on this example: the model-based $H_\infty$ strategy used in \cite{vollmer2001h} and the data-driven structured $H_\infty$ technique from \cite{apkarian2017structured}. Section \ref{lddc} then applies the \textbf{L-DDC} algorithm from \cite{kergus2019} on this application. The results are then given and compared with \cite{apkarian2017structured} in Section \ref{results}. Finally, concluding remarks, along with outlooks for future research, are proposed in Section \ref{conclusion}.

\section{The continuous crystallizer}
\label{sec:cry}

\subsection{Process description and control objective}
The considered application is the control of a continuous cooling crystallizer. It is a separation process widely used in the chemical industry. Its goal is to produce high-purity solids from liquids. The system has one input, the solute feed concentration $c_f(t)$, and one output, the solute concentration in the crystallizer $c(t)$. The state of the system is $x(t)=\left[n(L,t) \ c(t)\right]^T$, where $n(L,t)$ denotes the crystal size distribution. Physically, this system is described by population and mass balance equations. A complete mathematical model of this system is derived in \cite{rachah2016mathematical}.

The objective is to stabilize the plant around a desired steady-state $c(t)=c_{ss}=4.09 mol/L$, which is just above the saturation concentration $c_s=4.038mol/L$, required for the crystals to be produced. For this steady state, as said in \cite{vollmer2001h} and \cite{rachah2016mathematical}, the system is unstable and presents sustained oscillations which may degrade the quality of the crystals. Feedback control is therefore needed. When linearizing the system's partial differential equations around the desired steady state, the crystallizer is characterized by an irrational transfer function of infinite dimension:
\begin{equation}
    \color{black}{P(s)=\frac{\Delta c(s)}{\Delta c_f(s)}=\frac{p_{12}(s)}{p_{13}(s)+q_{12}(s)e^{-sk_f}+r_{12}(s)e^{-sk_p}},}
    \label{Pcry}
\end{equation}
where $p_{12}$, $q_{12}$, $r_{12}$ and $p_{13}$ are polynomials of order 12 and 13 respectively. The constants $k_f$ and $k_p$ depends on the fines and products removal sizes respectively, and on the crystal growth at steady-state.

The two approaches to control this system, presented in \cite{vollmer2001h} and \cite{apkarian2017structured} are recalled hereafter.

\subsection{Model-based $\mathcal{H}_\infty$ control design}
In \cite{vollmer2001h}, a weighted mixed sensitivity problem is formulated in order to make the controller robust to process variations (multiplicative uncertainty is considered) but also to disturbances. An infinite-dimensional $\mathcal{H}_\infty$ controller synthesis method from \cite{foias1996robust} is then applied to solve it using the irrational transfer function $P$ from \eqref{Pcry}, which makes this approach model-based. This procedure requires estimation of the plant's instabilities, which is done through direct search, in order to factorize the plant's expression. An irrational controller is finally obtained, which is then approximated by a rational transfer function of order 8 for implementation purposes.

\subsection{Data-driven structured $\mathcal{H}_\infty$ control design}
In \cite{apkarian2017structured}, a data-driven structured $\mathcal{H}_\infty$-control method for infinite-dimensional linear time-invariant systems is proposed and does not require use of any order reduction technique. A robust control problem is formulated and, based on a set of frequency-domain data, a nonsmooth trust-region bundle method is used and results in a locally optimal structured controller ensuring closed-loop exponential stability. During the optimization, a stability test is performed at each step thanks to a data-driven Nyquist test on the winding number. This test requires building a fine frequency grid on which data is extracted from the irrational model $P$. The procedure needs to be initialized with a stabilizing controller. In \cite{apkarian2017structured}, the continuous crystallizer is approximated by a rational model $P_{502}$ of order 502, obtained through a finite-difference method. This model $P_{502}$ is used to design an initial stabilizing controller. For the considered application, this method results in the following second-order controller $C$:
\begin{equation}
    C(s)=\frac{54.47s^2+2.317s+0.02446}{s^2+0.002033s+4.374\cdot 10^{-6}}.
    \label{K14}
\end{equation}

\section{Control design: the L-DDC approach}
\label{lddc}
The objective of this section is to tackle the control of the continuous crystallizer with the \textbf{L-DDC} algorithm \cite{kergus2019}. This technique allows the design of a reduced-order feedback controller that minimizes the difference between the closed-loop and the desired one, which is specified as a reference model $M$. 

Using a single set of frequency-domain data from the plant, it is possible to compute samples of the frequency response of the so-called ideal controller $K^\star$. It is defined as the controller that would give exactly the desired behaviour $M$ and, by definition:
\begin{equation}
    K^\star(\jmath\omega_{i})=P(\jmath \omega_{i})^{-1}M(\jmath\omega_{i}))(I_{n_y}-M(\jmath\omega_{i}))^{-1}.
    \label{Kideal}
\end{equation}
The main idea of the \textbf{L-DDC} is to formulate the design problem as approximation and reduction of the ideal controller $K^\star$ using the Loewner framework. The \textbf{L-DDC} procedure is composed of three steps: the choice of an achievable reference model is detailed in \ref{subsec:build_M} before moving to controller approximation and reduction in \ref{subsec:loewner} and a closed-loop stability test is finally considered in \ref{subsec:analysis}. In the present case, a frequency grid consists of $N=500$ frequencies, logspaced between $10^{-3}$ and 1 rad.s$^{-1}$ is considered. The corresponding samples of the frequency response of the plant are estimated directly through the irrational transfer function $P$ from \eqref{Pcry}, as in \cite{apkarian2017structured}.

\subsection{Step 1: Building achievable specifications}
\label{subsec:build_M}
Despite its apparent simplicity, the choice of the model-reference $M$ is a critical step, as underlined by \cite{bazanella2011data} and \cite{selvi2018towards} and \cite{kergus2019}. Indeed, the reference model should not only represent a desired closed-loop behaviour, but also respect the plant's fundamental limitations. As explained in \cite{havre2001achievable}, the performance limitations of a system depends on its instabilities and can be expressed as a set of interpolatory conditions on the closed-loop transfer function:
\begin{equation}
    \left\{\begin{array}{c}
        M(z_i) = 0  \\
        M(p_j) =1 
    \end{array}\right. .
    \label{eq:defM}
\end{equation}
where $\left\{z_i\right\}_{i=1}^{n_z}$ and $\left\{p_j\right\}_{j=1}^{n_p}$ are respectively the unstable zeros and poles of the plant. Considering that the inverse of the plant $P$ appears in $K^\star$ \eqref{Kideal}, it is clear that \eqref{eq:defM} prevents $K^\star$ from compensating the unstable poles and non-minimum phase zeros.

\begin{figure*}
    \centering
    \subfloat[][Projection of $\left\{P(\jmath \omega_i)\right\}_{i=1}^N$.]{\includegraphics[width=0.49\textwidth,trim=0cm 2cm 0cm 2cm, clip]{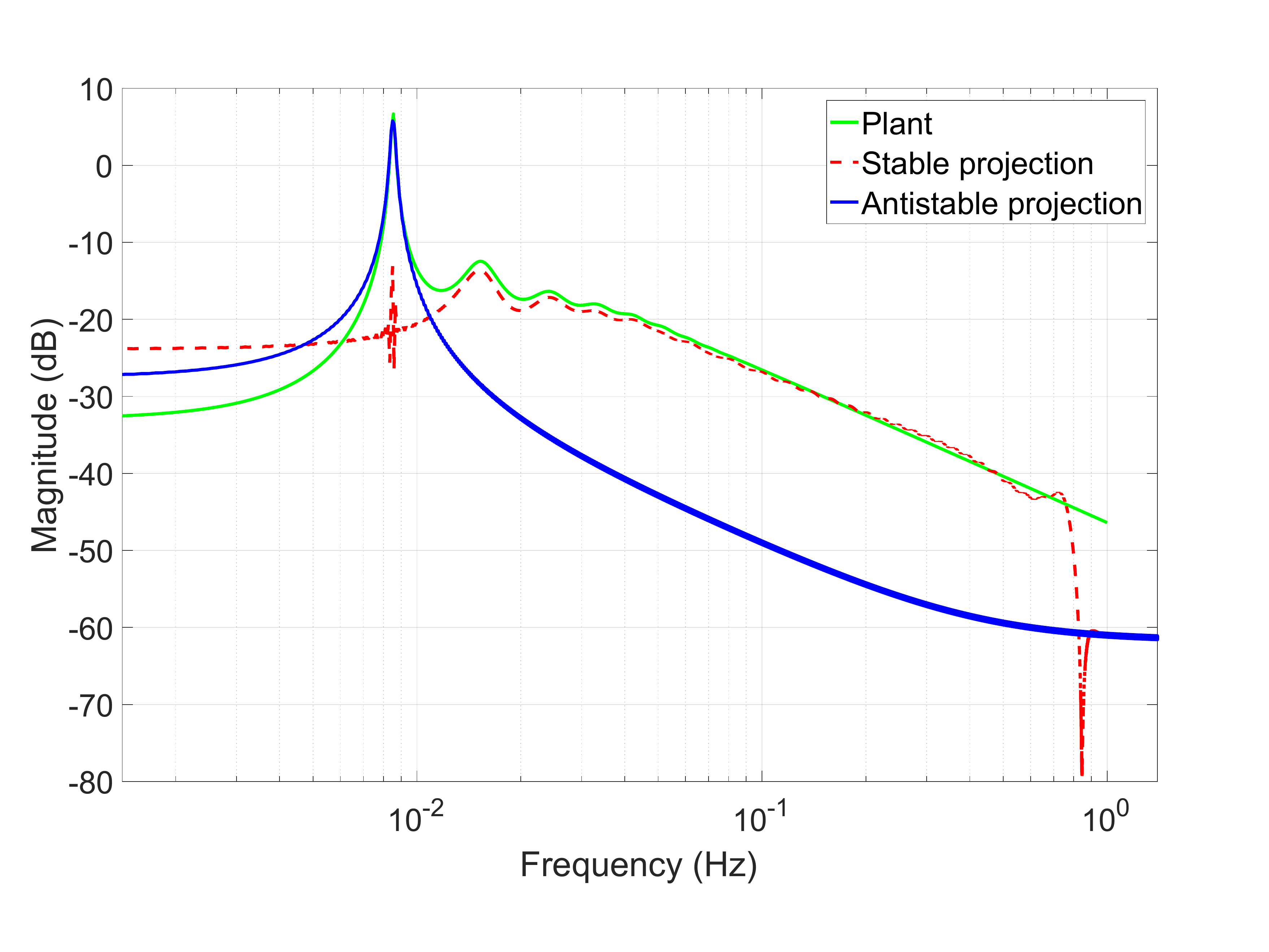}\label{proj_cry}}
    \subfloat[][Projection of $\left\{P^{-1}(\jmath \omega_i)\right\}_{i=1}^N$.]{\includegraphics[width=0.49\textwidth,trim=0cm 2cm 0cm 2cm, clip]{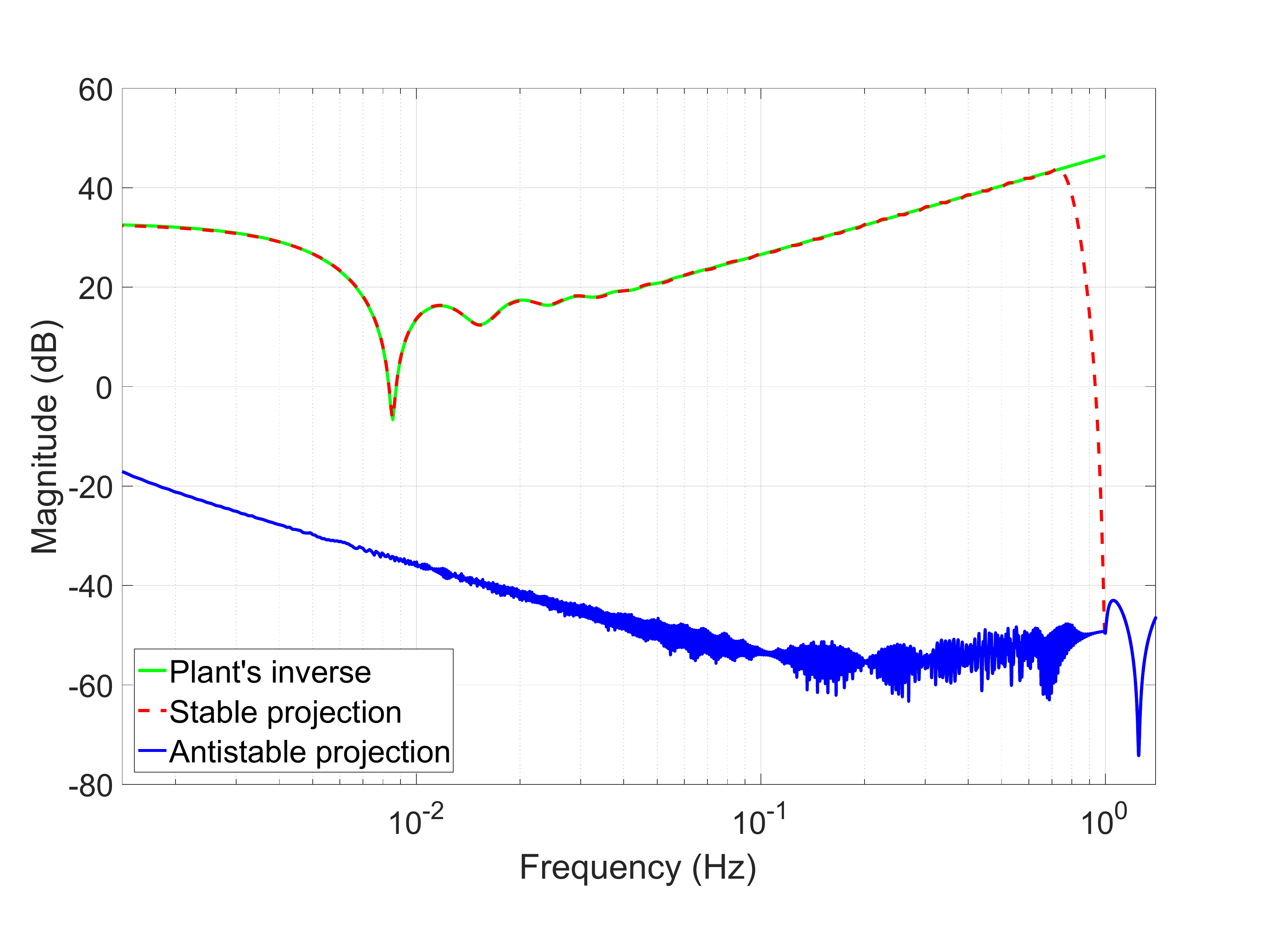}\label{proj_inv_cry}}
    \caption{Projection of the system's data: the system is unstable and minimum phase.}
    \label{proj_data}
\end{figure*}

It is therefore essential to consider the system's instabilities in order to define its performance limitations, which  may be hard to determine based on the available irrational transfer function $P$. To that extent, the frequency response of the plant, assumed to be in $\mathcal{L}_2$, is projected on the Hardy spaces $\mathcal{H}_2$ and $\overline{\mathcal{H}}_2$ as in \cite{cooman2018model}. For the continuous crystallizer and the considered frequency grid, the result is given on Figure \ref{proj_cry}: the antistable projection fits the resonance while the stable part fits the rest of the frequency-response of the plant. Therefore the plant is unstable. The  projection is then performed on $\left\{ \omega_i, P(\jmath \omega_i)^{-1}\right\}$, see Figure \ref{proj_inv_cry}: the stable projection of the plant's inverse fits the inverse of the plant's frequency-response samples, meaning that the plant is minimum phase.

\begin{remark}
In case $P\notin\mathcal{L}_2$, when $P$ has instabilities on the imaginary axis for instance, it is then filtered by a bandpass filter in order to be analyzed. The instabilities outside the filter's bandwidth cannot be detected, which constitutes a limitation of this approach.
\end{remark}

\begin{figure}
    \centering
    \includegraphics[width=0.49\textwidth,trim=0cm 0cm 0cm 2cm, clip]{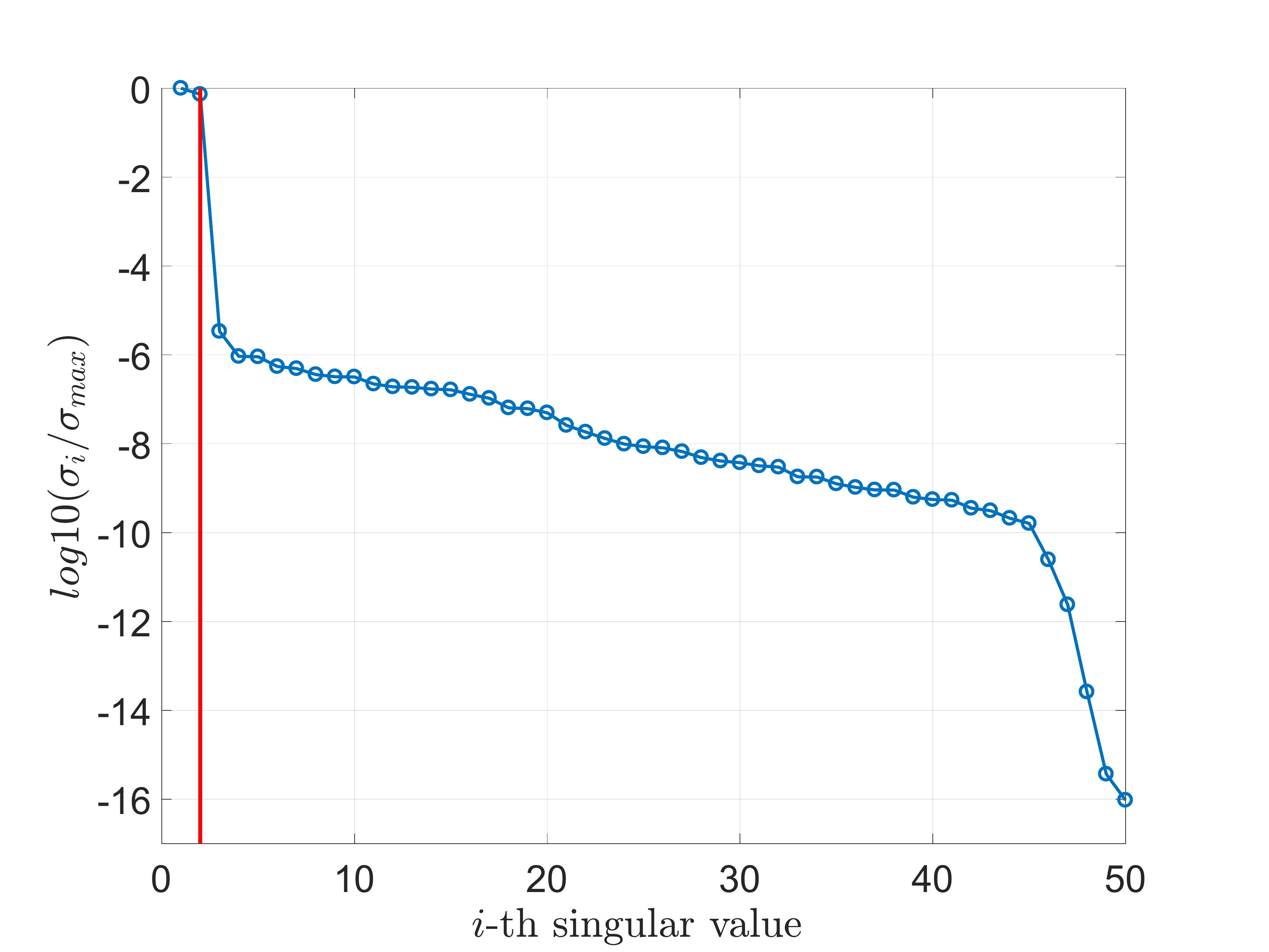}
    \caption{SVD of the Hankel matrix of the antistable projection of the plant's data: the system exhibits two RHP poles.}
    \label{svd_Hankel}
\end{figure}

Since the system is found to have unstable poles, they are estimated as in \cite{cooman2018estimating}. The Hankel matrix of the antistable projection of the plant's data is computed and a Singular Value Decomposition (SVD) is performed to reveal its rank, which corresponds to the number of unstable poles of the system $P$. The SVD is visible on Figure \ref{svd_Hankel}: according to the drop after the second singular value, the system exhibits two unstable poles.
These two Right-Half-Plane (RHP) poles are then estimated by reconstructing the observability matrix of the antistable projection. Their value is given in Table \ref{RHPpoles}. The obtained values are coherent with the ones found in \cite{vollmer2001h} and with the RHP poles of the rational model $P_{502}$ from \cite{apkarian2017structured}.

\begin{table}[H]
    \centering
    \begin{tabular}{|c|c|}
        \hline
        Estimated RHP poles & $1.07\times10^{-4}\pm 0.852\times 10^{-2}\jmath$ \\
        \hline
        RHP poles of $P_{502}$ & $3.83\times10^{-5}\pm 0.848\times 10^{-2}\jmath$ \\
        \hline
        Estimated RHP poles in \cite{vollmer2001h} & $0.99\times10^{-4}\pm 0.89\times 10^{-2}\jmath$\\
        \hline
    \end{tabular}
    \caption{Estimation of the RHP poles of the plant.}
    \label{RHPpoles}
    \vspace{-0.15cm}
\end{table}
\vspace{-0.2cm}

\begin{remark}
Estimating the plant's instabilities may seem like a model-based approach. However, estimating the plant's RHP poles and zeros is different from identifying a model of the system, which would require making the distinction between the approximation artefacts and the true instabilities of the system.
\end{remark}

The desired performances are given as initial reference model $M_{init}$ is a first order transfer function:
\begin{equation}
    M_{init}(s)=\frac{1}{1+\tau s}, \ \tau=1s.    
\end{equation}
The reference model is then made achievable as follows:
\begin{equation}
    M=1-(1-M_{init})B_p,
    \label{eq:Mf}
\end{equation}
with $B_p$ defined according to the estimated unstable poles of the plant from Table \ref{RHPpoles}, as follows:
\begin{equation}
    B_p(s)=\prod_{j=1}^{n_p}\frac{s-p_j}{s+p_j}.
    \label{blaschke_siso}
\end{equation}
Since $B(p_j)=0$, the reference model $M$ satisfies $M(p_j)=1$, i.e $M$ is achievable according to \eqref{eq:defM}. The filter $B_p$ is chosen because $\forall \omega, \ |B_p(\jmath\omega)|=1$: it allows not to modifiy the specified desired behaviour $M_{init}$ unnecessarily.

\subsection{Step 2: Controller interpolation and reduction}
\label{subsec:loewner}
\begin{figure*}
    \centering
    \subfloat[][Normalized SVD of the Loewner matrix.]{\includegraphics[width=0.49\textwidth,trim=0cm 1cm 0cm 2cm, clip]{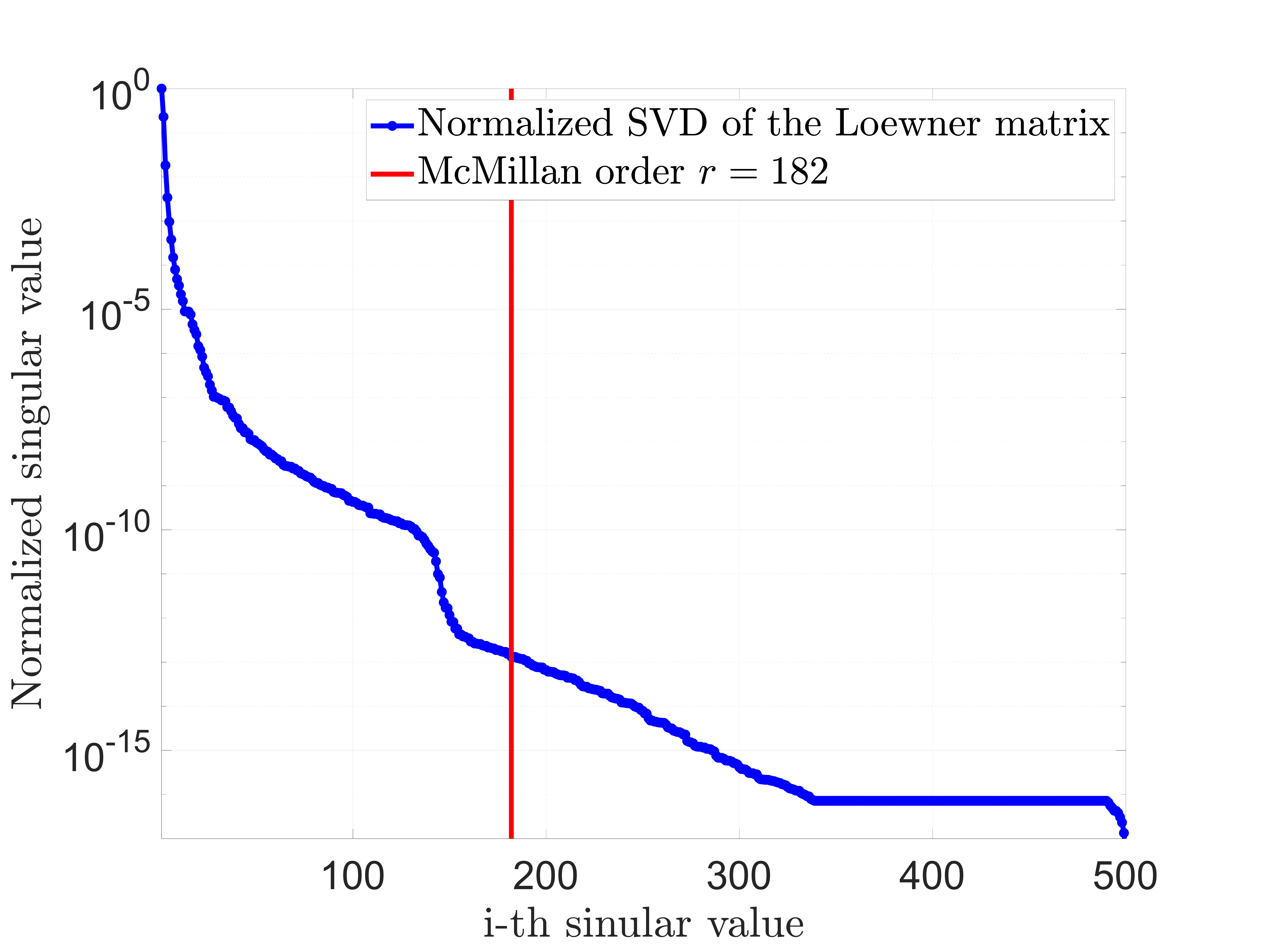}\label{svdL}}
    \subfloat[][Resulting controllers.]{\includegraphics[width=0.49\textwidth,trim=0cm 1cm 0cm 2cm, clip]{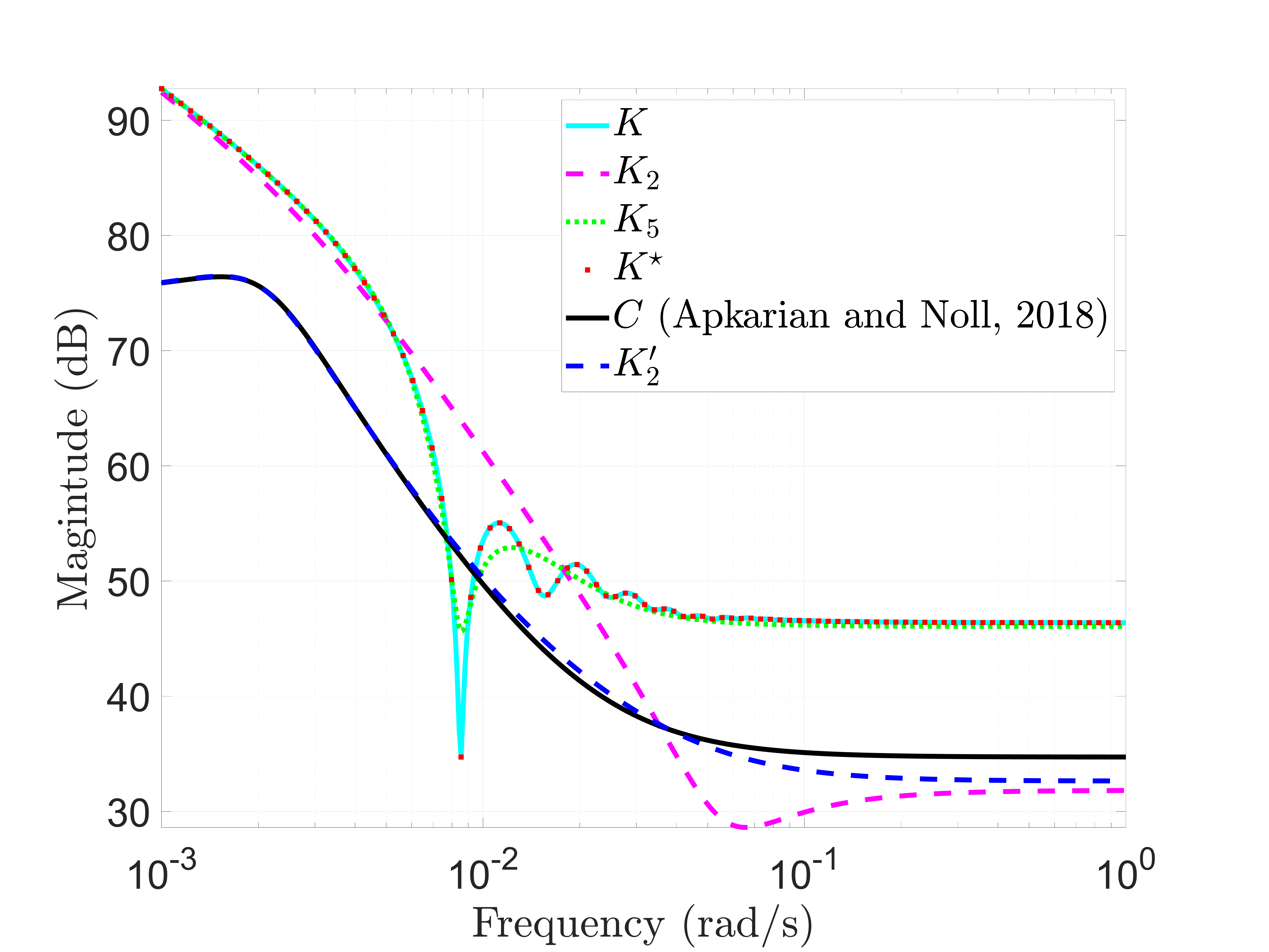}\label{idK}}
    \caption{Identification of the controller using the LDDC framework: using $M$ as a reference model, the minimal realization $K$ of the ideal controller $K^\star$ has 182 states and is stable, see the SVD on the left. It is reduced to a second-order controller $K_2$ and a fifth-order one $K_5$, whose frequency responses are given along with the one $C$ from \cite{apkarian2017structured} and a second-order controller $K_2'$ obtained by the \textbf{L-DDC} using another reference model $M'$, see paragraph \ref{subsec:influence_Minit}.}
    \label{controller}
\end{figure*}

Once an achievable reference model $M$ is obtained, the frequency response of the associated ideal controller is computed as in \eqref{Kideal}. The Loewner framework, detailed in \cite{antoulas2017tutorial}, \cite{mayo2007framework} and briefly recalled hereafter, is then used to obtain a minimal controller realization $K$ that interpolates the ideal controller frequency response samples:
\begin{equation}
    \forall i=1\dots N,\ K(\jmath \omega_i)=K^\star(\jmath \omega_i).
    \label{CI_K}
\end{equation}
In order to construct such a representation $K$, the sample points are divided into two sets of interpolation points: 
\begin{equation*}
    \left\{\jmath\omega_i\right\}_{i=1}^N=\left\{\mu_k\right\}_{k=1}^{N_\mu}\bigcup \left\{\lambda_j\right\}_{j=1}^{N_\lambda}
\end{equation*}
From these, Loewner and shifted-Loewner matrices can be built as
\begin{equation}
    \begin{array}{c}
         \left[\mathbb{L}\right]_{i,j}=\frac{ K^\star(\mu_i)- K^\star(\lambda_j)}{\mu_i-\lambda_j}\\
         
         \left[\mathbb{L}_\sigma\right]_{i,j}=\frac{\mu_i K^\star(\mu_i)-\lambda_j K^\star(\lambda_j)}{\mu_i-\lambda_j}  
    \end{array}.
\end{equation}

Then, a Singular Value Decomposition (SVD) is performed on the Loewner pencil $\left(\mathbb{L},\mathbb{L}_{\sigma}\right)$ such that,
\begin{equation}
    \begin{array}{lcr}
         \left[\mathbb{L},\mathbb{L}_{\sigma}\right]=Y_1\Sigma_1X_1^H&,&\left[\begin{array}{c}
                \mathbb{L}\\
               \mathbb{L}_\sigma
         \end{array}\right]=Y_2 \Sigma_2 X_2^H  \\
    \end{array},
    \label{svd_pencil}
\end{equation}
where $^H$ represents the conjugate transpose. During this process, the McMillan order $r$ of the interpolation model is computed as the number of non-zero singular values. Finally, a realization of the interpolation model is given in a descriptor form by:
\begin{equation}
    K:\left\{ \begin{array}{r@{=}l}
        E\dot{x} & Ax+Bu \\
        y & Cx
    \end{array}\right.
\end{equation}
where the matrices are defined as follows:
\small
\begin{equation}
         E=-Y_1^H\mathbb{L}X_2,\quad A=-Y_1^H\mathbb{L}_\sigma X_2,\quad B=Y_1^H V,\quad C=WX_2,
\end{equation}\normalsize
where $V_i^T= K^\star(\mu_i)$ and $W_j= K^\star(\lambda_j)$. A reduced order model of order $n$ can be obtained by keeping only the $n\leq r$ highest singular values in \eqref{svd_pencil}.

Figure \ref{svdL} exhibits the SVD which is used in the \textbf{L-DDC} approach to choose the controller order. In the present case, the obtained minimal realisation $K$ of the ideal controller $K^\star$ is of order 182. The controller is reduced to a second-order model to be compared with the one obtained in \cite{apkarian2017structured}, its expression is given in \eqref{K_cry}. The results are visible on Figure \ref{idK}. A 5th order controller $K_5$ is also represented to show how controller reduction affects the performance, see section \ref{results}.
\begin{equation}
    K_2(s)=\frac{39.082 (s^2 + 0.04164s + 0.003132)}{s (s+0.002751)}
    \label{K_cry}
\end{equation}

\subsection{Step 3: Stability analysis}
\label{subsec:analysis}
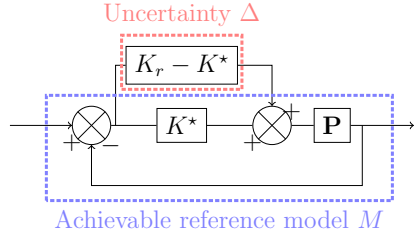
\begin{figure}[h]
    \centering
    \scalebox{0.8}{\tikzstyle{abstract}=[rectangle, draw=black, rounded corners, fill=blue!5, text centered, anchor=north, text=black, text width=3.5cm]
\tikzstyle{abstract2}=[rectangle, text centered, anchor=north, text=black]
\tikzstyle{block}=[rectangle, draw=black, text centered]
\tikzstyle{comment}=[rectangle, draw=black, rounded corners, fill=red!5, text centered, anchor=north, text=black, text width=3cm]
\tikzstyle{myarrow}=[->, >=open triangle 90, thick]
\tikzstyle{line}=[-, thick]
\tikzstyle{sum}=[draw,circle,text width = 0.3cm]
\begin{tikzpicture}
	\draw
	
	node at (0,0)[]{}
	node [name=ref] {} 
	node at (1.5,0)[sum](comp){}
	node at (3,0)[block](cor){$K^\star$}
	node at (3,1)[block](cor2){$K_r-K^\star$}
	node at (4.5,0)[sum](comp2){}
	node at (5.5,0)[block](plant){$\mathbf{P}$}
	node at (7,0)[](y1){};

	\draw (comp.135) -- (comp.315);
    \draw (comp.225) -- (comp.45);
    \node (plus) [below] at (comp.180) {$+$};
    \node (minus) [below] at (comp.350) {$-$};
    \draw (comp2.135) -- (comp2.315);
    \draw (comp2.225) -- (comp2.45);
    \node (plus) [below] at (comp2.180) {$+$};
    \node (minus) [above] at (comp2.350) {$+$};
	
	\draw[->](ref) -- node[below]{}(comp);
	\draw[->](comp) -- node[above]{}(cor) -- (comp2) -- node[above]{}(plant) --node[above]{}(y1);
	\draw[->](1.9,0) |- (cor2) -| (comp2);
	\draw[->](6,0) -- (6,-1) -| node[below]{}(comp);
	\draw[-, dash pattern=on 2pt off 1pt, line width=0.5mm, blue!50] (0.75,-1.25) -- (0.75,0.5) -- (6.5,0.5)--(6.5,-1.25)--node[below]{\color{blue!50}{Achievable reference model $M$}}(0.75,-1.25);
	\draw[-, dash pattern=on 2pt off 1pt, line width=0.5mm, red!50] (2,0.6) -- (4,0.6) -- (4,1.5)--node[above]{\color{red!50}{Uncertainty $\Delta$}}(2,1.5)--(2,0.6);
	\end{tikzpicture}}
    \caption{Stability analysis scheme: the controller modelling error appears as an uncertainty.}
    \label{fig:small_gain}
    \vspace{-0.2cm}
\end{figure}

While the choice of an achievable reference model $M$ guarantees that $K^\star$ stabilizes the plant, it is still needed to check if closed-loop stability is ensured for reduced order controllers. To that extent, as in \cite{van2009data}, the closed-loop obtained with a controller $K_r$ of order $r$ is written as the negative feedback interconnection between the controller modelling error $\Delta=K_r-K^\star$ and $P(1-M)$, as visible on Figure \ref{fig:small_gain}.
The choice of an achievable reference model $M$ and the stability of the controller allow us to apply the small-gain theorem as in \cite{van2009data} to derive the following internal stability condition: the interconnected system shown on Figure \ref{fig:small_gain} is well-posed and internally stable for all stable $\Delta=K_r-K^\star$ with $\left\Vert \Delta \right\Vert_\infty < \frac{1}{\gamma}$ if and only if $\left\Vert P(1-M) \right\Vert_\infty \leq\gamma$, where $\gamma>0$.

\begin{figure}[h]
    \centering
    \includegraphics[width=0.49\textwidth,trim=0cm 0cm 0cm 2cm, clip]{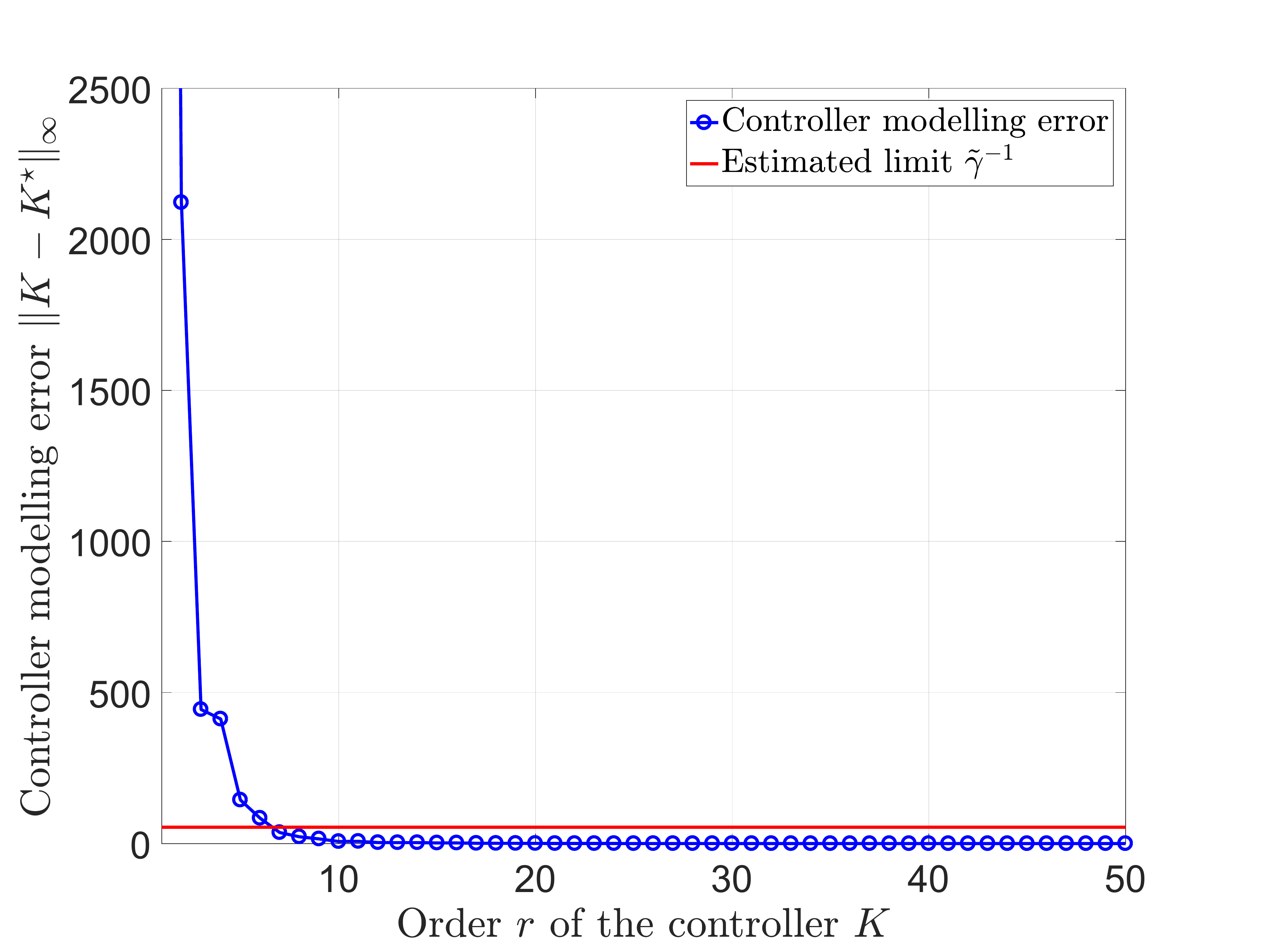}
    \caption{Stability analysis based on Theorem 1: the controller can be reduced up to an order 7 while preserving internal stability.}
    \label{fig:stab_analysis}
    \vspace{-0.2cm}
\end{figure}

It follows that limiting the controller modelling error, and therefore the controller reduction, ensures that internal stability is preserved. The limit requires to estimate $\gamma=\Vert P(1-M)\Vert_\infty$, which is done by taking $\Tilde{\gamma}=max|P(\jmath\omega_i)(1-M(\jmath\omega_i))|=54.2547$. This stability analysis is given on Figure \ref{fig:stab_analysis} and indicates that, for the continuous crystallizer, the ideal controller can be reduced up to an order 7 while preserving internal stability.
The main drawback of this approach is that the proposed stability test is very conservative. Indeed, it is not possible to conclude whether $K_2$ stabilizes the system or not, since $||K_2 - K^\star||_\infty=2121.8$. However, the closed-loop results, which will be discussed in the next section, indicate that $K_2$ stabilizes the plant internally.
\section{Results and discussions}
\label{results}
\begin{figure*}
    \centering
    \subfloat[][Resulting closed-loops.]{\includegraphics[width=0.49\textwidth,trim=0cm 1cm 0cm 2cm, clip]{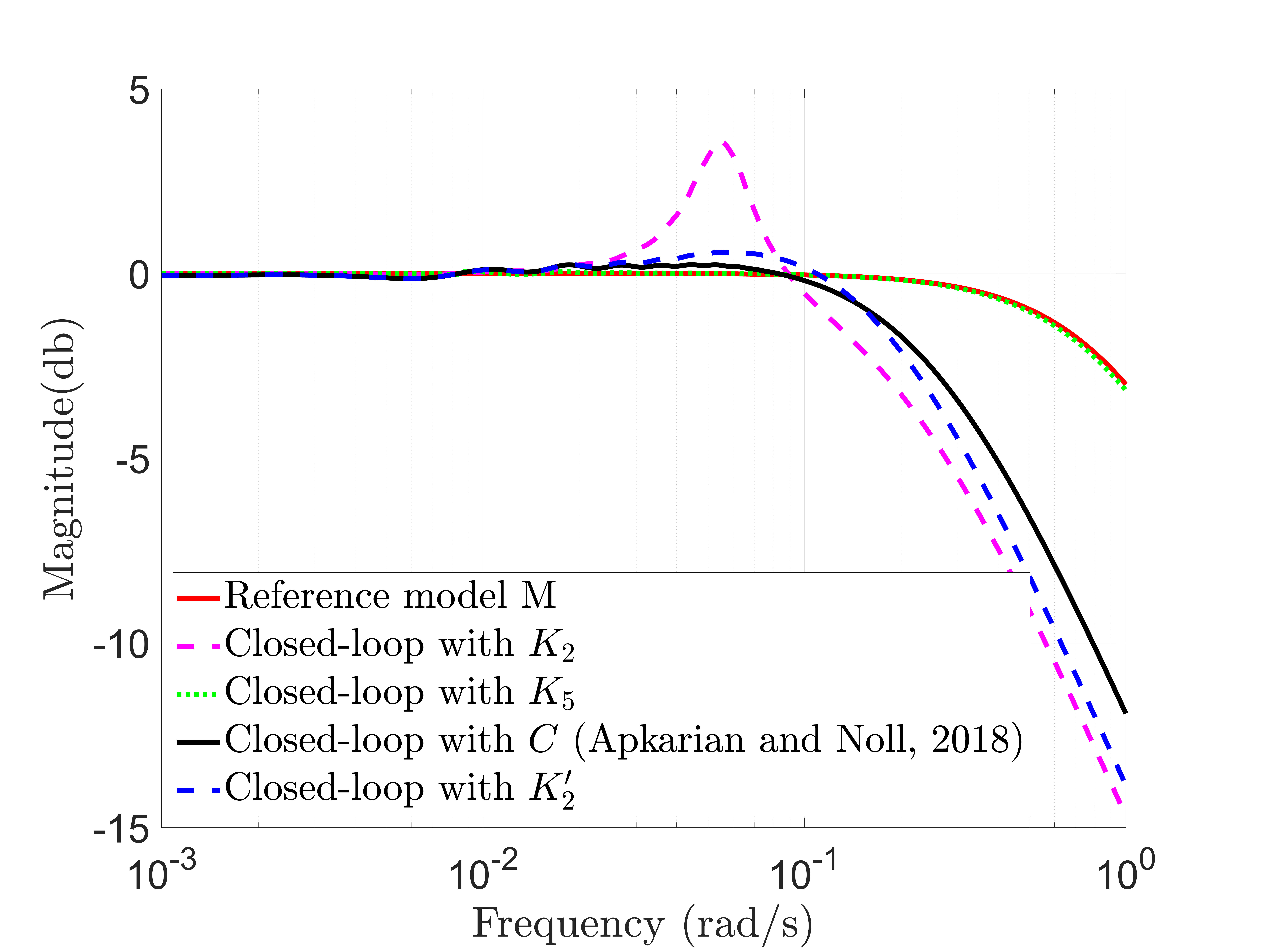}\label{closed_loop}}
    \subfloat[][Simulation of the passage to a new steady-state.]{\includegraphics[width=0.49\textwidth,trim=0cm 1cm 0cm 2cm, clip]{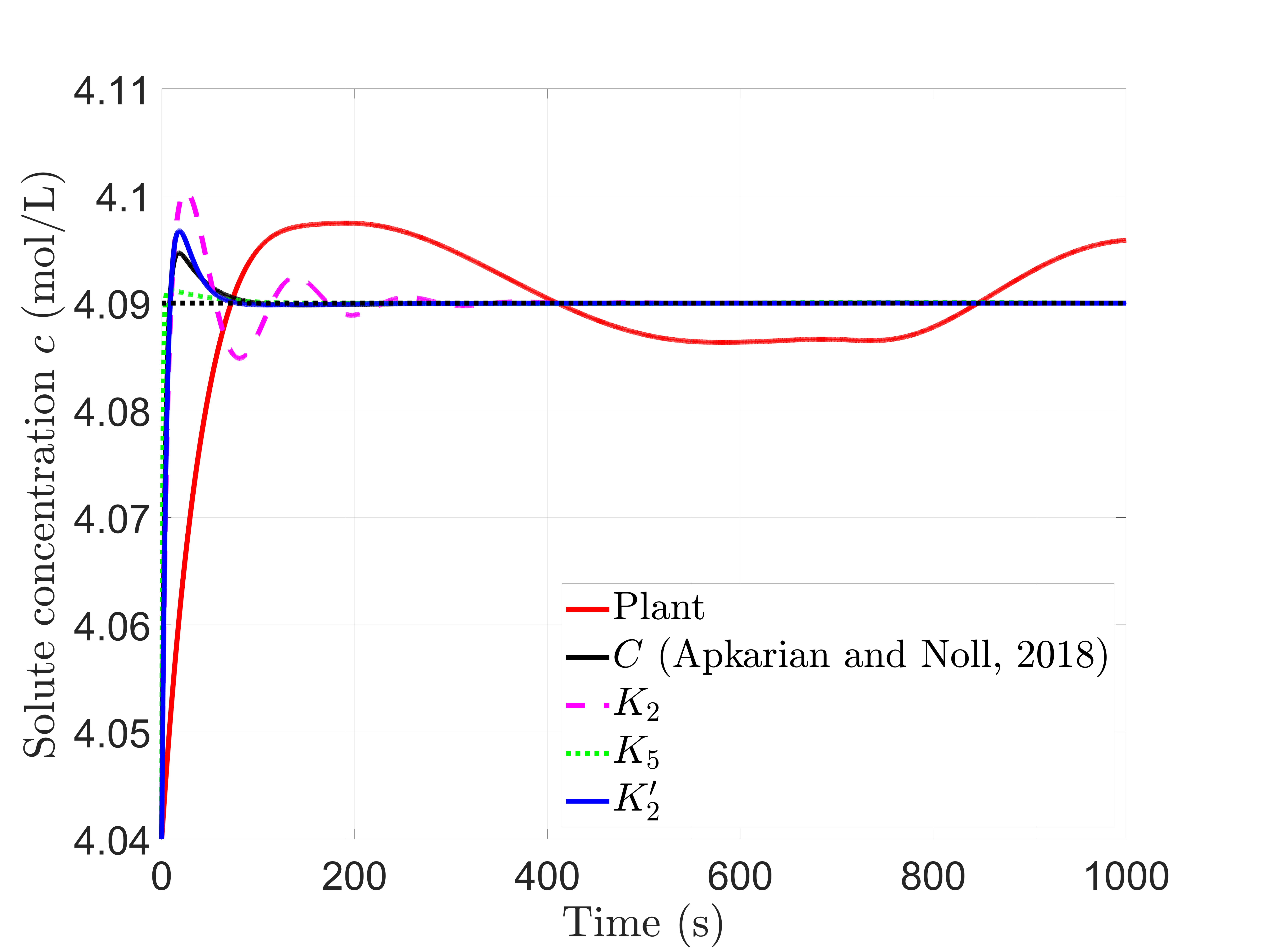}\label{simul_cry}}
    \caption{Closed-loop performances for the controllers $K_2$, $K_5$, $C$ from \cite{apkarian2017structured} and $K_2'$: closed-loop frequency response on the left and simulation of the passage to a new steady-state on the right.}
    \label{result}
    \vspace{-0.5cm}
\end{figure*}

\subsection{Closed-loop results}
The finite-difference model $P_{502}$ is used to simulate the closed-loop behaviour in time-domain. The identified controller $K_2$ is compared to $C$ from \cite{apkarian2017structured}. The results are visible on Figure \ref{simul_cry}. The design of $C$ in \cite{apkarian2017structured} allows reaching the desired steady-state faster with a smaller magnitude overshoot (10\% for $C$ and 20\% $K_2$). These better results might be because model-reference control is limited when it comes to the expression of the closed-loop specifications. Figure \ref{idK} highlights that $C$ does not compensate any mode of the system since it is the solution of a robust problem while the ideal controller and its minimal realization  $K$ reflect the stable dynamics of the plant due to the presence of $P^{-1}$ in the definition of $K^\star$ in \eqref{Kideal}. The results obtained with $K_2$ also underline that the reduction of the ideal controller degrades closed-loop control performance. As shown on Figure \ref{idK}, the higher the order of the identified controller, the more it will fit the frequency-response of the ideal controller. This is also visible on Figure \ref{closed_loop}, where the closed-loop transfer functions, built from \eqref{cl_estimate}, are represented. A high order controller is more likely to give the desired closed-loop behaviour, specified by the reference model $M$. In the present case, reducing the ideal controller to $K_2$ induces poor damping, as visible on Figure \ref{closed_loop}, explaining the oscillations on Figure \ref{simul_cry}. 

\subsection{Towards a less conservative stability test}
\begin{figure}[h]
    \centering
    \includegraphics[width=0.49\textwidth,trim=0cm 0cm 0cm 2cm, clip]{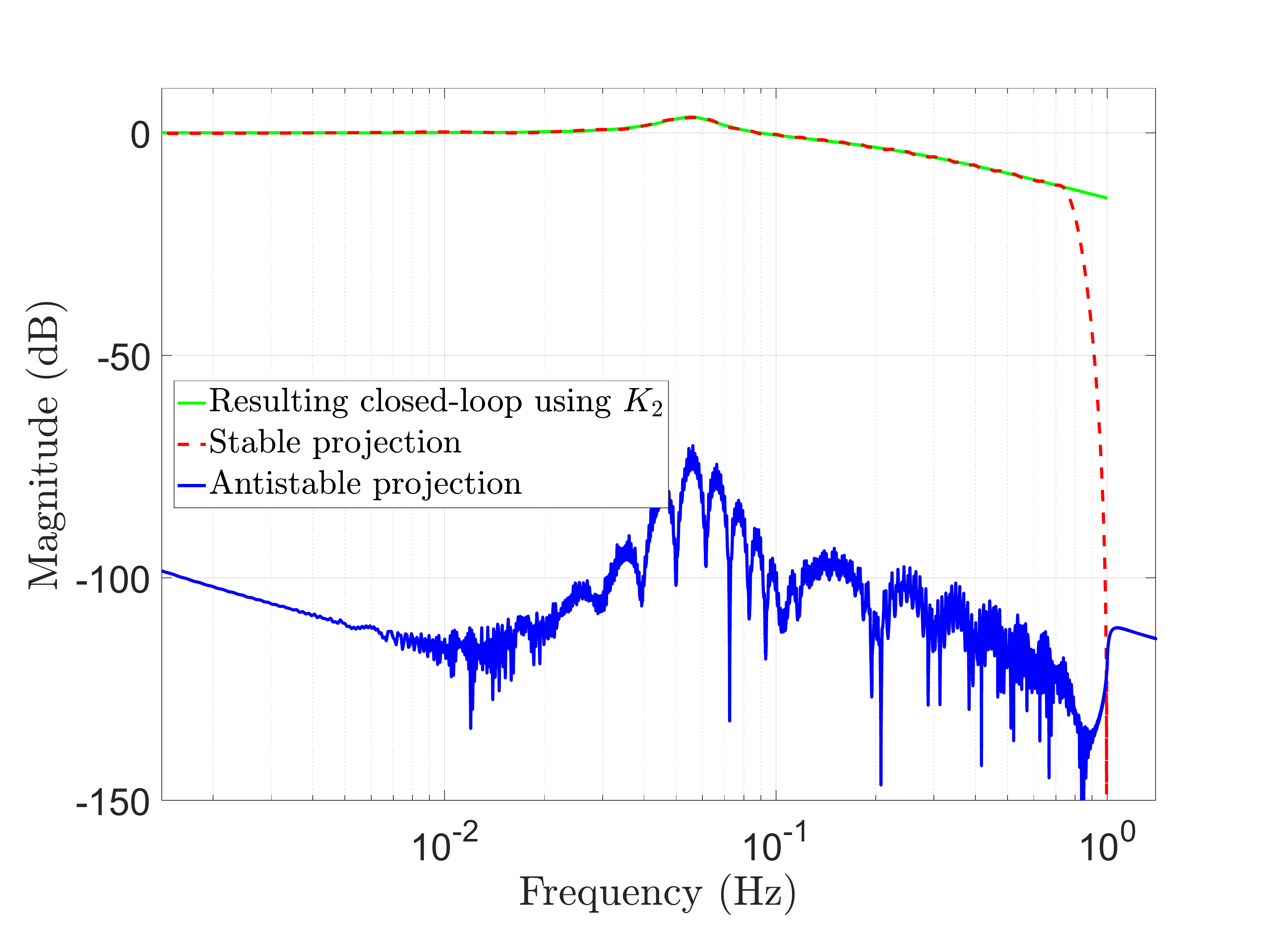}
    \caption{Projection of the closed-loop data reconstructed using the controller $K_2$: the input-output transfer function is concluded to be stable.}
    \label{alt_stab_analysis}
    \vspace{-0.6cm}
\end{figure}

First, the stability test from \ref{subsec:analysis} could not validate $K_2$ but the simulations on Figure \ref{result} suggest that $K_2$ is stabilizing. It is therefore important to provide another tool to analyse stability. One idea could be to use the projection from \cite{cooman2018model}, previously used to detect the system's instabilities in \ref{subsec:build_M}, on the reconstructed closed-loop data $\left\{ H(\jmath \omega_i)\right\}_{i=1}^N$:
\begin{equation}
    H(\jmath \omega_i)=(I+P(\jmath \omega_i)K(\jmath \omega_i))^{-1}P(\jmath \omega_i)K(\jmath \omega_i).
    \label{cl_estimate}
\end{equation}
The result is visible on Figure \ref{alt_stab_analysis}. From there, it is possible to conclude that the input-output transfer function is stable. Since the instabilities of the plant have been estimated, it is known that they are not compensated by $K_2$, which therefore stabilizes the plant internally.

\subsection{Influence of the initial reference model}
\label{subsec:influence_Minit}
Finally, the compromise between controller's complexity and closed-loop performance does not only depend on the controller reduction but also on the initial reference model specified by the user. In \cite{apkarian2017structured}, the specifications are given as frequency weighting functions, giving more freedom to the desired closed-loop behaviour. To underline this aspect, the closed-loop reached by the controller $C$ obtained in \cite{apkarian2017structured}, denoted $M'$, is taken as reference model. $M'$ has been computed using base on \eqref{cl_estimate}. It leads to the identification of the second-order controller $K_2'$:
\begin{equation}
    K_2'(s)=\frac{27.578 (s^2 + 0.06562s + 0.004418)}{(s+1.026e-06) (s+0.002737)}.
    \label{KM14}
\end{equation}
$K_2'$ gives a better response time and a bigger magnitude overshoot than $C$. However, the closed-loop performances induced by $K_2'$ are much closer to the ones obtained in \cite{apkarian2017structured} than the ones obtained by $K_2$ when using $M$ as a reference model.

\subsection{Discussion}

First, it is interesting to note that in the three control techniques mentioned in this paper, \cite{vollmer2001h}, \cite{apkarian2017structured} and \cite{kergus2019}, the system's fundamental limitations play a key role in the success of the design. These limitations can be determined by estimating the plant's instabilities as in \cite{vollmer2001h} or \cite{kergus2019} or by assuming that an initial stabilizing controller is available, as in \cite{apkarian2017structured}.

Indeed, in \cite{apkarian2017structured}, the initial stabilizing controller enables a stability test on the winding number. However, this method requires to build a fine frequency grid ($N=10^5$ in \cite{apkarian2017structured} against $N=500$ here) on which samples of the frequency-response of the plant are assumed to be available. Furthermore, the control design relies on iterative non-smooth optimization, which can be time-consuming and is sensitive to the considered initial stabilizing controller. On the other side, the main strength of the \textbf{L-DDC} algorithm is its simplicity. It is a one shot technique, free of strong assumptions, and does not depend on an initial stabilizing controller. 

However, the main advantage of the approach from \cite{apkarian2017structured} is that it guarantees stability and performances, while the \textbf{L-DDC} is only able to ensure closed-loop stability. As a matter of fact, the projection-based stability test suggested in \ref{alt_stab_analysis} remains a boolean test and the controller reduction step does not take into account robustness and closed-loop performances. 

In addition, the results detailed in this section show that choosing one desired transfer function, even if it respects the plant's fundamental limitations, is a limitation of the \textbf{L-DDC} approach, compared to robust specifications using frequency weightings as in \cite{apkarian2017structured}. Consequently, if the obtained performances are not sufficient for the user, it may be hard to improve the results by changing the reference model without any guideline. Nonetheless, it is possible to increase the order of the controller (if that is available) to get closer to the ideal case.

Regarding model-based strategies, using the infinite-dimensional model for controller synthesis as in \cite{vollmer2001h} results in an irrational optimal controller, which is then approximated by a reduced 8th order rational controller. The impact of the controller reduction step is not studied in \cite{vollmer2001h}. Another model-based strategy could be to apply a $\mathcal{H}_\infty$ structured technique, \cite{apkarian2006nonsmooth} for instance, on a reduced-order rational model $P_r$ of the system, obtained through the Loewner framework for example. However, the obtained controller would then be tailored to $P_r$ and not to the real system. Contrary to these model-based strategies, the \textbf{L-DDC} approach offer ways to make sure that the actual closed-loop, composed of the reduced controller and the real system, is stable. 

Finally, it should be noted that the \textbf{L-DDC} and the model-based strategies all require the use of reduction when dealing with an infinite dimensional system such as the continuous crystallizer. They then face a common challenge: how can closed-loop performance guide the choice of the complexity-accuracy trade-off during model or controller reduction?

\section{Conclusions}
\label{conclusion}
In this paper, the control of a continuous crystallizer, described by an infinite dimensional transfer function, illustrates the efficiency of the data-driven approaches from \cite{apkarian2017structured} and \cite{kergus2019} (\textbf{L-DDC}), on this category of problems. Indeed, they allow to obtain a reduced-order controller tailored to the actual system from samples of the frequency response of the system, which can be evaluated easily and directly from irrational transfer functions. This is a strong advantage over the model-based design proposed in \cite{vollmer2001h} that results in an irrational controller that needs to be reduced to be implemented.

The results and the comparison between these different techniques indicate that there is a need to formalize how controller reduction affects the closed-loop performances and robustness, in the data-driven case as seen in the \textbf{L-DDC} approach as much as in the model-based approach of \cite{vollmer2001h}. In addition, further work should investigate the possibility to tune the reference model to have a better control on the corresponding closed-loop performances. A first option would be to design the reference model not only to respect the plant's fundamental limitations but also to correspond to a certain level of robustness.




\bibliographystyle{unsrt}
\bibliography{biblio}

\begin{thebibliography}{10}

\bibitem{hou2013model}
Z-S. Hou and Z.~Wang.
\newblock From model-based control to data-driven control: Survey,
  classification and perspective.
\newblock {\em Information Sciences}, 2013.

\bibitem{foias1996robust}
C.~Foias, H.~{\"O}zbay, and A.~Tannenbaum.
\newblock {\em Robust control of infinite dimensional systems: frequency domain
  methods}.
\newblock Springer, 1996.

\bibitem{morris2010control}
K.~Morris and WS~Levine.
\newblock Control of systems governed by partial differential equations.
\newblock {\em The control theory handbook}, 2010.

\bibitem{kergus2019}
P.~{Kergus}, M.~{Olivi}, C.~{Poussot-Vassal}, and F.~{Demourant}.
\newblock From reference model selection to controller validation: Application
  to loewner data-driven control.
\newblock {\em IEEE Control Systems Letters}, 2019.

\bibitem{apkarian2017structured}
P.~Apkarian and D.~Noll.
\newblock {Structured $H_\infty$ control of infinite dimensional systems}.
\newblock {\em International Journal of Robust and Nonlinear Control}, 2018.

\bibitem{rachah2016mathematical}
A.~Rachah, D.~Noll, F.~Espitalier, and F.~Baillon.
\newblock A mathematical model for continuous crystallization.
\newblock {\em Mathematical Methods in the Applied Sciences}, 2016.

\bibitem{vollmer2001h}
U.~Vollmer and J.~Raisch.
\newblock {$H_\infty$-Control of a continuous crystallizer}.
\newblock {\em Control Engineering Practice}, 2001.

\bibitem{bazanella2011data}
AS. Bazanella, L.~Campestrini, and D.~Eckhard.
\newblock {\em Data-driven controller design: the H2 approach}.
\newblock Springer, 2011.

\bibitem{selvi2018towards}
D.~Selvi, D.~Piga, and A.~Bemporad.
\newblock Towards direct data-driven model-free design of optimal controllers.
\newblock In {\em European Control Conference}. IEEE, 2018.

\bibitem{havre2001achievable}
K.~Havre and S.~Skogestad.
\newblock Achievable performance of multivariable systems with unstable zeros
  and poles.
\newblock {\em International Journal of Control}, 2001.

\bibitem{cooman2018model}
A.~Cooman, F.~Seyfert, M.~Olivi, S.~Chevillard, and L.~Baratchart.
\newblock Model-free closed-loop stability analysis: A linear functional
  approach.
\newblock {\em IEEE Trans. on Microwave Theory and Techniques}, 2018.

\bibitem{cooman2018estimating}
A.~Cooman, F.~Seyfert, and S.~Amari.
\newblock Estimating unstable poles in simulations of microwave circuits.
\newblock In {\em IEEE/MTT-S International Microwave Symposium}, 2018.

\bibitem{antoulas2017tutorial}
Athanasios~C Antoulas, Sanda Lefteriu, A~Cosmin Ionita, P~Benner, and A~Cohen.
\newblock A tutorial introduction to the loewner framework for model reduction.
\newblock {\em Model Reduction and Approximation: Theory and Algorithms}, 2017.

\bibitem{mayo2007framework}
A.~J. Mayo and A.~C. Antoulas.
\newblock A framework for the solution of the generalized realization problem.
\newblock {\em Linear algebra and its applications}, 2007.

\bibitem{van2009data}
K.~van Heusden, A.~Karimi, and D.~Bonvin.
\newblock Data-driven controller validation.
\newblock In {\em 15th IFAC Symposium on System Identification}, 2009.

\bibitem{apkarian2006nonsmooth}
P~Apkarian and D~Noll.
\newblock Nonsmooth h-infinity synthesis.
\newblock {\em IEEE Trans. on Automatic Control}, 2006.

\end{thebibliography}


\end{document}